\documentclass[aip,jcp,reprint,floatfix,showkeys]{revtex4-1}
\usepackage{graphicx}
\usepackage{color}
\usepackage{amsmath}
\usepackage{amssymb}
\usepackage{dcolumn}
\usepackage{bm}
\newcommand{\un}[1]{\ensuremath{\unskip\,\mathrm{#1}}}

\DeclareGraphicsExtensions{.eps}

\begin{document}

\title{Short-range dynamics of a nematic liquid-crystalline phase}

\author{Andreas S. Poulos}
\author{Doru Constantin}
\email{constantin@lps.u-psud.fr}
\author{Patrick Davidson}
\author{Brigitte Pansu}
\affiliation{Laboratoire de Physique des Solides, Universit\'{e}
Paris-Sud, CNRS, UMR 8502, 91405 Orsay, France.}

\author{\'{E}ric Freyssingeas}
\affiliation{Laboratoire de Physique, \'Ecole Normale Sup\'erieure
de Lyon, CNRS, UMR 5672, 69364 Lyon, France.}

\author{Anders Madsen}
\affiliation{European Synchrotron Radiation Facility, Bo\^{\i}te
Postale 220, 38043 Grenoble, France.}

\author{Corinne Chan\'{e}ac}
\affiliation{Laboratoire de Chimie de la Mati\`{e}re Condens\'{e}e,
Universit\'{e} Paris 6, CNRS, UMR 7574, 75252 Paris, France.}

\date{\today}

\begin{abstract}
Using X-ray photon correlation spectroscopy, we studied the dynamics
in the nematic phase of a nanorod suspension. The collective
diffusion coefficient in the plane perpendicular to the director
varies sharply with the wave vector. Combining the structure factor
and the diffusion coefficient we find that the hydrodynamic function
of the phase decreases by more than a factor of ten when going from
length scales comparable to the inter-particle distance towards
larger values. Thus, the collective dynamics of the nematic phase
experiences strong and scale-dependent slowing down, in contrast
with isotropic suspensions of slender rods or of spherical
particles.

\end{abstract}




\maketitle


Nematics are the simplest example of a phase with no positional
order, but still exhibiting orientational order (and hence
anisotropy). This combination endows them with remarkable qualities:
although fluid, they have elastic properties and, consequently,
long-lived fluctuations. A great deal is known about the large-scale
behaviour of nematic systems, which is well-described by a
generalized hydrodynamic model \cite{Martin72}. This theoretical
description was confirmed (and refined) using a wealth of
experimental techniques. The method of choice for studying
nematodynamics is dynamic light scattering (DLS), which is sensitive
to the relaxation of nematic fluctuations on micron scales
\cite{Orsay69,Straley74}.

On the other hand, there is much less data on the short-range
dynamics of the nematic phase, convering length scales comparable to the
inter-particle distance. In this limit, the continuous medium model
is bound to break down, and more microscopic considerations must be
taken into account. Since this is the scale at which interaction
between particles defines the structure of the system, understanding
the dynamics is essential for building a complete picture of the
phase. A considerable body of theoretical and numerical work exists
\cite{Kirchhoff96,Jose06}, but there is very little experimental
data, mainly due to the lack of adapted techniques (due to the
typical particle size, this range of scattering vectors is out of
reach for DLS.) Alternative methods can be used, such as inelastic
neutron scattering, which is however limited to sub-microsecond
dynamics (too fast for cooperative processes). Spin relaxation has
also been employed, but it lacks the required space resolution and
the conclusions are indirect.

X-ray scattering techniques are suitable for exploring these
distances, but until recently they were only able to draw a
\textit{static} picture of the system. This situation is changing
due to the progress of X-ray photon correlation spectroscopy (XPCS),
opening up the time dimension. However, due to inherent technical
difficulties, the experimental systems must fulfill very stringent
conditions, such as slow relaxation rates and high scattering
contrast. Some experiments have already been performed on nematics
using XPCS, but they were only concerned with very slow relaxation
in a gel phase \cite{Bandyopadhyay04} or with capillary surface
waves \cite{Madsen03}.

Essential information on the physics of multi-particle systems is
contained in the structure factor $S(q)$ and the collective
diffusion coefficient $D(q)$ \cite{Pusey91}. Based on very general thermodynamic
arguments, these parameters are related by $D(q) \sim 1/S(q)$ (``de
Gennes narrowing'' \cite{deGennes59}). In colloidal suspensions, a
more refined treatment must take into account the hydrodynamic
interactions, which further modulate this dependence.

In this letter, we study a fluid nematic phase of goethite
($\alpha$-FeOOH) nanorods (with moderate aspect ratio) and determine
the hydrodynamic function over a $q$-range corresponding to length
scales comparable to the inter-particle distances. Unexpectedly, we
find that for wave vectors $q < q_{\mathrm{max}}$ (below the maximum
of the structure factor), the dynamics of the system slows down
considerably. This result is in stark contrast with isotropic
suspensions of slender rods, where no hydrodynamic effect is
observed \cite{Graf91}. Furthermore, the effect is much stronger
than the variation of the hydrodynamic function in suspensions of
colloidal spheres \cite{Robert08}, emphasizing the role of particle
anisotropy and of the nematic order.

\begin{figure} [htbp]
\centering
\includegraphics[width=8.4cm, keepaspectratio=true]{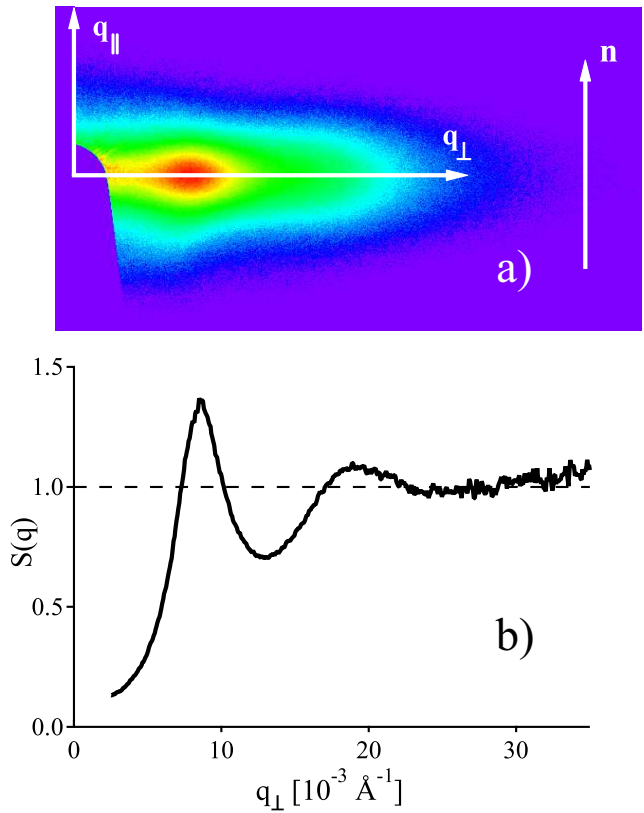}
\caption{a) SAXS pattern of an aligned nematic sample with
$\phi_{g}$ = 6.7\% at 228~K. The arrow indicates the direction of the
nematic director $\vec{n}$. b) Structure factor along the $q_{\bot}$
direction calculated from the SAXS data in a).}
\label{figure1_structure}
\end{figure}

Goethite nanoparticles were synthesized according to a well-established
procedure and dispersed in water \cite{Thies07,Lemaire04_1}. The particles are lath-shaped; transmission electron microscopy images show that they are $4000 \pm 1000$~{\AA} long and $330 \pm
110$~{\AA} wide (mean and standard deviation obtained from a
log-normal fit). Moreover, the particle cross-section is usually anisotropic by a factor of 2.5 \cite{Lemaire04_1}. In the following, we therefore consider that the particles rotate freely around their main axis, as expected in a uniaxial nematic phase. In
order to slow down their dynamics so that it may be conveniently
studied by XPCS, the particles were resuspended in an 80~wt.\%
propane-1,3-diol in water mixture. Although this mixture has lower
dielectric permittivity than water, the particles still bear some
positive electrical surface charge density, which ensures the
colloidal stability of the suspension. In the first approximation,
the effect of electrostatic interactions can be described by
introducing an effective radius larger than the bare one, which
tends to reduce the apparent particle aspect ratio \cite{Vroege92}.
For example, previous studies of another goethite suspension in pure
water have shown that the ratio of effective diameter to bare
diameter is about $1.6$ \cite{Lemaire04_2}. With the solvent mixture
used here, this electrostatic correction factor is expected to be
smaller. Therefore, in the following, the particles will be
considered as cylindrical rods, 4000~{\AA} long and with an aspect
ratio of about 10. (The interpretation presented below does not
critically depend on the precise value of the effective aspect
ratio.) At volume fractions $\phi_{g} \leq 4.2\%$ the suspensions
are isotropic, whereas at $\phi_{g} \geq 6.7\%$ they form a nematic
liquid-crystalline phase that aligns in low magnetic field, with its
nematic director parallel to the field \cite{Lemaire02}. Samples of
different volume fractions were held in optical flat glass
capillaries, 50~$\mu$m thick (VitroCom, NJ, USA) and placed in a vacuum chamber. The
temperature was lowered to 228~K, where the viscosity of the solvent
is of the order of 1000~mPa.s. The nematic phase was aligned with a
150~mT field that was then removed for the actual measurements.

The small-angle X-ray scattering (SAXS) and XPCS measurements were
performed at the TROIKA beam line ID10A of the ESRF with an X-ray
energy of 8~keV ($\lambda=1.55$~\AA) selected by a single-bounce
Si(111) monochromator, in the uniform filling mode of the storage
ring. A (partially) coherent beam is obtained by inserting a 10
$\mu$m pinhole aperture a few centimetres upstream of the sample.

\begin{figure} [htbp]
\includegraphics[width=8.4cm, keepaspectratio=true]{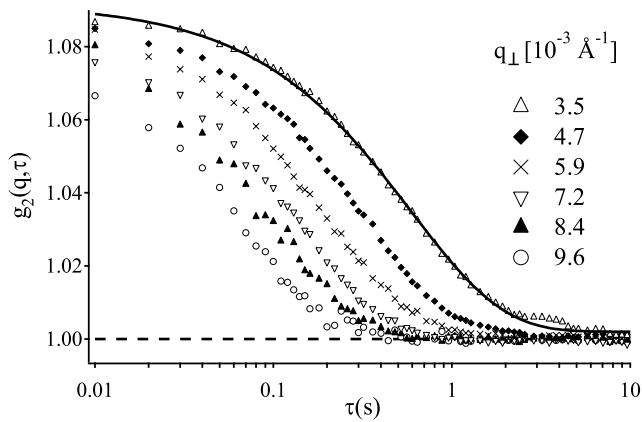}
\caption{Autocorrelation functions for a nematic sample with
$\phi_{g}$ = 6.7\% at different scattering vectors $q_{\bot}$. The
solid line is the fit with a stretched exponential (the stretching
exponent is about 0.6 for all the curves).} \label{acf_medipix}
\end{figure}

For the XPCS measurements, we used a two-dimensional (2D) Maxipix detector consisting
of $256 \times 256$ square pixels (55~$\mu$m in size), and the
intensity autocorrelation functions were calculated by ensemble
averaging over equivalent pixels \cite{Fluerasu07}. In the nematic
phase, the pixels averaged were restricted to a narrow slice
perpendicular to the nematic director for q$_{\bot}$ (see Figure \ref{figure1_structure} for an illustration and Figure
\ref{acf_medipix} for the results), or parallel to the nematic
director for q$_{\|}$. In the isotropic phase, all pixels at the
same scattering vector modulus were averaged. Some measurements were
also performed using a point detector (an avalanche photodiode)
connected to an external digital correlator (Flex01D-08).

A typical SAXS pattern of an aligned nematic sample is shown in
Figure \ref{figure1_structure}a. The static structure factor
$S(q_{\bot})$ in the $q_{\bot}$ direction (Figure
\ref{figure1_structure}b) was obtained by dividing the scattered
intensity by the form factor measured independently on a dilute
solution. $S(q_{\bot})$ displays a well-defined interaction peak, at
a value $q_{\mathrm{max}} = 8.6 \times 10^{-3}~\un{\AA}^{-1}$  due
to the liquid-like positional short-range order of the nanorods in
the plane perpendicular to the director.

\begin{figure}[htbp]
\includegraphics[width=8.4cm, keepaspectratio=true]{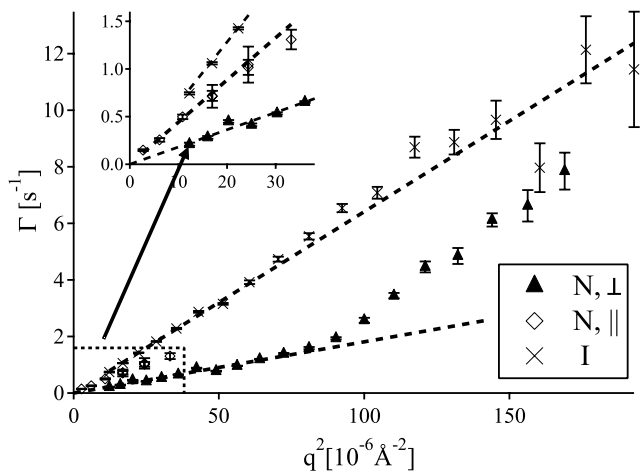}
\caption{Dispersion relations for a nematic ($\phi_{g}$ = 6.7\%)
sample, with $\vec{q}$ parallel to the nematic director
($\diamond$), and with $\vec{q}$ perpendicular to the nematic
director ($\blacktriangle$). The dispersion relation of an
isotropic, ($\phi_{g}$ = 2\%) sample is shown for comparison
($\times$).} \label{figure3_dispersion}
\end{figure}

The various dispersion relations are shown in Figure
\ref{figure3_dispersion}. The relaxation rate in the nematic phase,
perpendicular to the director ($N\, , \bot$, solid triangles) is
linear in $q_{\bot}^2$, roughly up to the position of the structure
peak. At higher $q_{\bot}$, the slope increases abruptly, before
approaching a final linear regime. In contrast, the behaviour along
the director ($N\, , \|$, open diamonds) is linear over the
accessible range, which is limited by the rapid fall-off of the
intensity in this direction (see Figure \ref{figure1_structure}a).
In the isotropic phase (with $\phi_{g}$ = 2\%), the dispersion
relation is also linear over the entire range.

\begin{figure}[htbp]
\includegraphics[width=8.4cm, keepaspectratio=true]{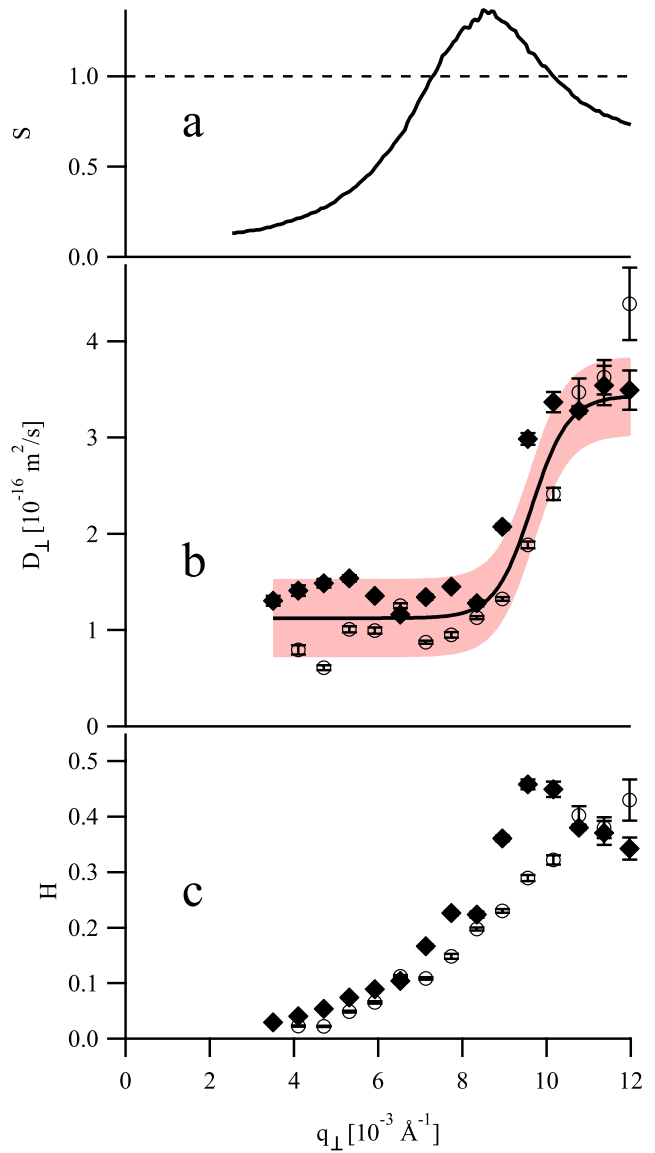}
\caption{a) Static structure factor $S(q_{\bot})$. b) Diffusion
coefficient $D_{N,\bot}(q_{\bot})$. The symbols correspond to two
different measurements performed on the same sample. The solid line
and the greyed area are the fit by a sigmoidal function and the $\pm
\sigma$ prediction bands (see text). c) Hydrodynamic function
obtained using Equation \ref{hydro_1}. The data are plotted against
a common $q_{\bot}$ axis.} \label{figure4_hydrodynamic}
\end{figure}

Let us define the collective diffusion coefficient $D_{N,\bot}(
q_{\bot}) = \Gamma (q_{\bot}) / q_{\bot}^2$, shown in Figure
\ref{figure4_hydrodynamic}b). For reference, panel a) shows the
structure factor (Figure \ref{figure1_structure}b).

Two regimes (above and below the interaction peak of the structure
factor) can be clearly distinguished. Between them, the diffusion
coefficient jumps by more than a factor of 3. To quantify this
variation, we fit $D_{N,\bot}( q_{\bot})$ with a sigmoidal function
(solid line in Figure \ref{figure4_hydrodynamic}b):
\begin{equation}
\label{sigmoidal} D(q) = D_{\mathrm{min}} + \frac{D_{\mathrm{max}} -
D_{\mathrm{min}}}{1+ \exp \left ( - \frac{q - q_{1/2}}{\Delta q}
\right )}
\end{equation}
with parameters: $D_{\mathrm{min}} = 1.1 \times 10^{-16}$ and
$D_{\mathrm{max}} = 3.4 \times 10^{-16} \un{m^2/s}$ while $q_{1/2}
= 9.6 \times 10^{-3}$ and $\Delta q = 0.45 \times 10^{-3}
\un{\AA^{-1}}$. The greyed areas around the fit are the $\pm \sigma$
prediction bands, quantifying the data scatter (about 68~\% of the
experimental points should fall within this area). The fit function
(\ref{sigmoidal}) is only chosen for convenience; there is no
physical reason for adopting it.

The decrease of $D_{N,\bot}( q_{\bot})$ at low wave vectors is even
more striking if we recall that, for $q_{\bot}$ vectors below the
peak, where the structure factor decreases, the diffusion
coefficient should \textit{increase}! We can restate this result
more precisely in terms of the \textit{hydrodynamic function},
$H(q_{\bot})$, defined by \cite{Naegele97}:
\begin{equation}
H(q_{\bot}) = \frac{D_{N,\bot} (q_{\bot})}{D_{0,\bot}} S(q_{\bot})
\label{hydro_1}
\end{equation}
and shown in Figure \ref{figure4_hydrodynamic}c).

Care must be exercised when determining the constant factor
$D_{0,\bot}$, which is the diffusion coefficient of the particles at
infinite dilution. The $\bot$ subscript specifies that one should
consider the diffusion in the direction perpendicular to the main
axis of the particle (in the nematic phase this axis is parallel to
the director, and we are now concerned with diffusion perpendicular
to the director). We start by measuring the (orientationally
averaged) diffusion coefficient in the isotropic phase, $D_I$. For
$\phi_{g}$ = 2\%, the data is shown in Figure
\ref{figure3_dispersion} (crosses), along with a linear fit yielding
$D_I(2 \%) = 6.4 \times 10^{-16}\un{m}^{2}\un{s}^{-1}$. Similar
measurements at $\phi_{g} = 0.5\%$ (data not shown) give $D_I(0.5
\%) = 8.0 \times 10^{-16}\un{m}^{2}\un{s}^{-1}$. Correcting these
results for the intrinsic viscosity of nanorod suspensions
\cite{Solomon98} leads to an infinite dilution value $D_0 = D_I(0) =
8.5 \times 10^{-16}\un{m}^{2}\un{s}^{-1}$. In isotropic solution,
$D_0$ is a geometric combination of the diffusion coefficients along
and perpendicular to the major axis: $3 D_0 = D_{0,\|} + 2
D_{0,\bot}$ and, for nanorods with an aspect ratio of 10,
$D_{0,\bot}/D_{0} = 0.88$ \cite{Tirado84}, resulting in $D_{0, \bot}
= 7.5 \times 10^{-16}\un{m}^{2}\un{s}^{-1}$.

At high $q_{\bot}$, $H(q)$ reaches a value lower than one
($H(\infty) \simeq 0.4$). For spherical particles, this limit is
described by \cite{Beenakker84}: $H(\infty) = \eta _0 / \eta$, where
$\eta$ is the viscosity of the suspension and $\eta _0$ that of the
solvent. Transposing this formula to our (non-spherical) system
yields fairly good agreement \footnote{Considering an ionic strength
of 10~mM, the data of Solomon and Boger \cite{Solomon98} give a
value of $\eta _0 / \eta = 0.41$ using the low concentration
expansion and 0.25 using the overall formula (for an aspect ratio of
8.4). Rheology measurements performed at room temperature in aqueous
suspension at the same concentration (6.7~\%) also yield $\eta _0 /
\eta = 0.25$, but the comparison is not straightforward due to the
lack of information on the electrostatic effects in the presence of
propane-1,3-diol.}. The main feature of $H(q)$ is however the
decrease at lower wave vector, starting around $q_{\mathrm{max}}$
and clearly visible in Figure \ref{figure4_hydrodynamic}c). At the
lowest accessible wave vector $H(q) \simeq 0.04$, ten times smaller
than the maximum value $H(\infty)$.

Hydrodynamic slowing down of the collective relaxation is also
encountered in suspensions of spherical particles, but its amplitude
is much lower; for a volume fraction $\phi = 9 \%$ (higher than in
our nematic phase), $H(0)/H(\infty) \simeq 1/3$ \cite{Robert08}.
Experimental and theoretical results for even higher volume
fractions of spheres (both in the low- and high-salt concentration
regimes) yield more modest decreases for the collective diffusion
\cite{Banchio06}. We conclude that the behaviour of $H(q)$ at low
wave vectors in the nematic phase is very different from that in
sphere suspensions.

A possible explanation for this difference is that, if the major
axis of the particles is much longer than the typical distance
between them, (in the limit of very large order parameter and aspect ratio)the nematic phase should behave like a two-dimensional
(2D) system \footnote{We define a 2D system as invariant under
translation along the normal direction, in contrast with the
extensively studied quasi-2D systems that consist of particles
confined at an interface or between rigid boundaries.}, where
hydrodynamic interactions are stronger than for a three-dimensional
system \cite{Happel83}. This explanation is in qualitative agreement
with recent simulations of 2D colloidal suspensions, where the
hydrodynamic interactions slow down the collective diffusion
coefficient \cite{Falck04}. In particular, for the volume fraction
used in our study these authors find that the hydrodynamic
interactions slow down the collective diffusion by at least a factor
of four (Ref. 26, Figure 2b). However, more theoretical or
numerical results would be needed, in particular concerning the
length-scale dependence of the diffusion coefficient $D(q)$, to
understand this pronounced slowing down.

\bigskip

\begin{acknowledgments}
A.S.Poulos gratefully acknowledges support from a Marie Curie action
(MEST-CT-2004-514307) and from a Triangle de la Physique contract
(OTP 26784).
\end{acknowledgments}


\begin{thebibliography}{10}%
\makeatletter
\providecommand \@ifxundefined [1]{%
 \ifx #1\undefined \expandafter \@firstoftwo
 \else \expandafter \@secondoftwo
\fi
}%
\providecommand \@ifnum [1]{%
 \ifnum #1\expandafter \@firstoftwo
 \else \expandafter \@secondoftwo
\fi
}%
\providecommand \enquote [1]{``#1''}%
\providecommand \bibnamefont  [1]{#1}%
\providecommand \bibfnamefont [1]{#1}%
\providecommand \citenamefont [1]{#1}%
\providecommand\href[0]{\@sanitize\@href}%
\providecommand\@href[1]{\endgroup\@@startlink{#1}\endgroup\@@href}%
\providecommand\@@href[1]{#1\@@endlink}%
\providecommand \@sanitize [0]{\begingroup\catcode`\&12\catcode`\#12\relax}%
\@ifxundefined \pdfoutput {\@firstoftwo}{%
 \@ifnum{\z@=\pdfoutput}{\@firstoftwo}{\@secondoftwo}%
}{%
 \providecommand\@@startlink[1]{\leavevmode}%
 \providecommand\@@endlink[0]{}%
}{%
 \providecommand\@@startlink[1]{%
  \leavevmode
  \pdfstartlink
   attr{/Border[0 0 1 ]/H/I/C[0 1 1]}%
   user{/Subtype/Link/A<</Type/Action/S/URI/URI(#1)>>}%
  \relax
 }%
 \providecommand\@@endlink[0]{\pdfendlink}%
}%
\providecommand \url  [0]{\begingroup\@sanitize \@url }%
\providecommand \@url [1]{\endgroup\@href {#1}{\urlprefix}}%
\providecommand \urlprefix [0]{URL }%
\providecommand \Eprint[0]{\href }%
\@ifxundefined \urlstyle {%
  \providecommand \doi [1]{doi:\discretionary{}{}{}#1}%
}{%
  \providecommand \doi [0]{doi:\discretionary{}{}{}\begingroup
  \urlstyle{rm}\Url }%
}%
\providecommand \doibase [0]{http://dx.doi.org/}%
\providecommand \Doi[1]{\href{\doibase#1}}%
\providecommand \selectlanguage [0]{\@gobble}%
\providecommand \bibinfo [0]{\@secondoftwo}%
\providecommand \bibfield [0]{\@secondoftwo}%
\providecommand \translation [1]{[#1]}%
\providecommand \BibitemOpen[0]{}%
\providecommand \bibitemStop [0]{}%
\providecommand \bibitemNoStop [0]{.\EOS\space}%
\providecommand \EOS [0]{\spacefactor3000\relax}%
\providecommand \BibitemShut [1]{\csname bibitem#1\endcsname}%
\bibitem{Martin72}%
  \BibitemOpen
  \bibfield{author}{%
  \bibinfo {author} {\bibfnamefont{P.~C.}\ \bibnamefont{Martin}}, \bibinfo
  {author} {\bibfnamefont{O.}~\bibnamefont{Parodi}},\ and\ \bibinfo {author}
  {\bibfnamefont{P.~S.}\ \bibnamefont{Pershan}},\ }%
  \bibfield{journal}{%
  \bibinfo {journal} {Physical Review A}\ }%
  \textbf{\bibinfo {volume} {6}},\ \bibinfo {pages} {2401} (\bibinfo {year}
  {1972})\BibitemShut{NoStop}%
\bibitem{Orsay69}%
  \BibitemOpen
  \bibfield{author}{%
  \bibinfo {author} {\bibnamefont{{Orsay Liquid Crystal Group}}},\ }%
  \bibfield{journal}{%
  \bibinfo {journal} {Physical Review Letters}\ }%
  \textbf{\bibinfo {volume} {22}},\ \bibinfo {pages} {1361} (\bibinfo {year}
  {1969})\BibitemShut{NoStop}%
\bibitem{Straley74}%
  \BibitemOpen
  \bibfield{author}{%
  \bibinfo {author} {\bibfnamefont{M.~J.}\ \bibnamefont{Stephen}}\ and\
  \bibinfo {author} {\bibfnamefont{J.~P.}\ \bibnamefont{Straley}},\ }%
  \bibfield{journal}{%
  \bibinfo {journal} {Rev. Mod. Phys.}\ }%
  \textbf{\bibinfo {volume} {46}},\ \bibinfo {pages} {617} (\bibinfo {year}
  {1974})\BibitemShut{NoStop}%
\bibitem{Kirchhoff96}%
  \BibitemOpen
  \bibfield{author}{%
  \bibinfo {author} {\bibfnamefont{T.}~\bibnamefont{Kirchhoff}}, \bibinfo
  {author} {\bibfnamefont{H.}~\bibnamefont{L\"{o}wen}},\ and\ \bibinfo {author}
  {\bibfnamefont{R.}~\bibnamefont{Klein}},\ }%
  \bibfield{journal}{%
  \bibinfo {journal} {Physical Review E}\ }%
  \textbf{\bibinfo {volume} {53}},\ \bibinfo {pages} {5011} (\bibinfo {year}
  {1996})\BibitemShut{NoStop}%
\bibitem{Jose06}%
  \BibitemOpen
  \bibfield{author}{%
  \bibinfo {author} {\bibfnamefont{P.~P.}\ \bibnamefont{Jose}}\ and\ \bibinfo
  {author} {\bibfnamefont{B.}~\bibnamefont{Bagchi}},\ }%
  \bibfield{journal}{%
  \bibinfo {journal} {J. Chem. Phys.}\ }%
  \textbf{\bibinfo {volume} {125}},\ \bibinfo {pages} {184901} (\bibinfo {year}
  {2006})\BibitemShut{NoStop}%
\bibitem{Bandyopadhyay04}%
  \BibitemOpen
  \bibfield{author}{%
  \bibinfo {author} {\bibfnamefont{R.}~\bibnamefont{Bandyopadhyay}}, \bibinfo
  {author} {\bibfnamefont{D.}~\bibnamefont{Liang}}, \bibinfo {author}
  {\bibfnamefont{H.}~\bibnamefont{Yardimci}}, \bibinfo {author}
  {\bibfnamefont{D.~A.}\ \bibnamefont{Sessoms}}, \bibinfo {author}
  {\bibfnamefont{M.~A.}\ \bibnamefont{Borthwick}}, \bibinfo {author}
  {\bibfnamefont{S.~G.~J.}\ \bibnamefont{Mochrie}}, \bibinfo {author}
  {\bibfnamefont{J.~L.}\ \bibnamefont{Harden}},\ and\ \bibinfo {author}
  {\bibfnamefont{R.~L.}\ \bibnamefont{Leheny}},\ }%
  \bibfield{journal}{%
  \bibinfo {journal} {Physical Review Letters}\ }%
  \textbf{\bibinfo {volume} {93}},\ \bibinfo {pages} {228302} (\bibinfo {year}
  {2004})\BibitemShut{NoStop}%
\bibitem{Madsen03}%
  \BibitemOpen
  \bibfield{author}{%
  \bibinfo {author} {\bibfnamefont{A.}~\bibnamefont{Madsen}}, \bibinfo {author}
  {\bibfnamefont{J.}~\bibnamefont{Als-Nielsen}},\ and\ \bibinfo {author}
  {\bibfnamefont{G.}~\bibnamefont{Gr\"{u}bel}},\ }%
  \bibfield{journal}{%
  \bibinfo {journal} {Physical Review Letters}\ }%
  \textbf{\bibinfo {volume} {90}},\ \bibinfo {pages} {085701} (\bibinfo {year}
  {2003})\BibitemShut{NoStop}%
\bibitem{Pusey91}%
  \BibitemOpen
  \bibfield{author}{%
  \bibinfo {author} {\bibfnamefont{P.~N.}\ \bibnamefont{Pusey}},\ }%
  in\ \emph{\bibinfo {booktitle} {Liquids, Freezing and the Glass Transition}}\
  (\bibinfo {publisher} {North-Holland},\ \bibinfo {address} {Amsterdam},\
  \bibinfo {year} {1991})\ pp.\ \bibinfo {pages} {763--942},\ \bibinfo {note}
  {section 5.2}\BibitemShut{NoStop}%
\bibitem{deGennes59}%
  \BibitemOpen
  \bibfield{author}{%
  \bibinfo {author} {\bibfnamefont{P.~G.}\ \bibnamefont{de~Gennes}},\ }%
  \bibfield{journal}{%
  \bibinfo {journal} {Physica}\ }%
  \textbf{\bibinfo {volume} {25}},\ \bibinfo {pages} {825} (\bibinfo {year}
  {1959})\BibitemShut{NoStop}%
\bibitem{Graf91}%
  \BibitemOpen
  \bibfield{author}{%
  \bibinfo {author} {\bibfnamefont{C.}~\bibnamefont{Graf}}, \bibinfo {author}
  {\bibfnamefont{M.}~\bibnamefont{Deggelmann}}, \bibinfo {author}
  {\bibfnamefont{M.}~\bibnamefont{Hagenb\"{u}chle}}, \bibinfo {author}
  {\bibfnamefont{H.}~\bibnamefont{Kramer}}, \bibinfo {author}
  {\bibfnamefont{R.}~\bibnamefont{Krause}}, \bibinfo {author}
  {\bibfnamefont{C.}~\bibnamefont{Martin}},\ and\ \bibinfo {author}
  {\bibfnamefont{R.}~\bibnamefont{Weber}},\ }%
  \bibfield{journal}{%
  \bibinfo {journal} {J. Chem. Phys.}\ }%
  \textbf{\bibinfo {volume} {95}},\ \bibinfo {pages} {6284} (\bibinfo {year}
  {1991})\BibitemShut{NoStop}%
\bibitem{Robert08}%
  \BibitemOpen
  \bibfield{author}{%
  \bibinfo {author} {\bibfnamefont{A.}~\bibnamefont{Robert}}, \bibinfo {author}
  {\bibfnamefont{J.}~\bibnamefont{Wagner}}, \bibinfo {author}
  {\bibfnamefont{W.}~\bibnamefont{H{\"a}rtl}}, \bibinfo {author}
  {\bibfnamefont{T.}~\bibnamefont{Autenrieth}},\ and\ \bibinfo {author}
  {\bibfnamefont{G.}~\bibnamefont{Gr{\"u}bel}},\ }%
  \bibfield{journal}{%
  \bibinfo {journal} {The European Physical Journal E-Soft Matter}\ }%
  \textbf{\bibinfo {volume} {25}},\ \bibinfo {pages} {77} (\bibinfo {year}
  {2008})\BibitemShut{NoStop}%
\bibitem{Thies07}%
  \BibitemOpen
  \bibfield{author}{%
  \bibinfo {author} {\bibfnamefont{D.~M.~E.}\ \bibnamefont{Thies-Weesie}},
  \bibinfo {author} {\bibfnamefont{J.~P.}\ \bibnamefont{de~Hoog}}, \bibinfo
  {author} {\bibfnamefont{M.~H.~H.}\ \bibnamefont{Mendiola}}, \bibinfo {author}
  {\bibfnamefont{A.~V.}\ \bibnamefont{Petukhov}},\ and\ \bibinfo {author}
  {\bibfnamefont{G.~J.}\ \bibnamefont{Vroege}},\ }%
  \bibfield{journal}{%
  \bibinfo {journal} {Chem. Mater}\ }%
  \textbf{\bibinfo {volume} {19}},\ \bibinfo {pages} {5538} (\bibinfo {year}
  {2007})\BibitemShut{NoStop}%
\bibitem{Lemaire04_1}%
  \BibitemOpen
  \bibfield{author}{%
  \bibinfo {author} {\bibfnamefont{B.~J.}\ \bibnamefont{Lemaire}}, \bibinfo
  {author} {\bibfnamefont{P.}~\bibnamefont{Davidson}}, \bibinfo {author}
  {\bibfnamefont{J.}~\bibnamefont{Ferr{\'e}}}, \bibinfo {author}
  {\bibfnamefont{J.~P.}\ \bibnamefont{Jamet}}, \bibinfo {author}
  {\bibfnamefont{D.}~\bibnamefont{Petermann}}, \bibinfo {author}
  {\bibfnamefont{P.}~\bibnamefont{Panine}}, \bibinfo {author}
  {\bibfnamefont{I.}~\bibnamefont{Dozov}},\ and\ \bibinfo {author}
  {\bibfnamefont{J.~P.}\ \bibnamefont{Jolivet}},\ }%
  \bibfield{journal}{%
  \bibinfo {journal} {The European Physical Journal E-Soft Matter}\ }%
  \textbf{\bibinfo {volume} {13}},\ \bibinfo {pages} {291} (\bibinfo {year}
  {2004})\BibitemShut{NoStop}%
\bibitem{Vroege92}%
  \BibitemOpen
  \bibfield{author}{%
  \bibinfo {author} {\bibfnamefont{G.~J.}\ \bibnamefont{Vroege}}\ and\ \bibinfo
  {author} {\bibfnamefont{H.~N.~W.}\ \bibnamefont{Lekkerkerker}},\ }%
  \bibfield{journal}{%
  \bibinfo {journal} {Reports on Progress in Physics}\ }%
  \textbf{\bibinfo {volume} {55}},\ \bibinfo {pages} {1241} (\bibinfo {year}
  {1992})\BibitemShut{NoStop}%
\bibitem{Lemaire04_2}%
  \BibitemOpen
  \bibfield{author}{%
  \bibinfo {author} {\bibfnamefont{B.~J.}\ \bibnamefont{Lemaire}}, \bibinfo
  {author} {\bibfnamefont{P.}~\bibnamefont{Davidson}}, \bibinfo {author}
  {\bibfnamefont{D.}~\bibnamefont{Petermann}}, \bibinfo {author}
  {\bibfnamefont{P.}~\bibnamefont{Panine}}, \bibinfo {author}
  {\bibfnamefont{I.}~\bibnamefont{Dozov}}, \bibinfo {author}
  {\bibfnamefont{D.}~\bibnamefont{Stoenescu}},\ and\ \bibinfo {author}
  {\bibfnamefont{J.~P.}\ \bibnamefont{Jolivet}},\ }%
  \bibfield{journal}{%
  \bibinfo {journal} {The European Physical Journal E-Soft Matter}\ }%
  \textbf{\bibinfo {volume} {13}},\ \bibinfo {pages} {309} (\bibinfo {year}
  {2004})\BibitemShut{NoStop}%
\bibitem{Lemaire02}%
  \BibitemOpen
  \bibfield{author}{%
  \bibinfo {author} {\bibfnamefont{B.~J.}\ \bibnamefont{Lemaire}}, \bibinfo
  {author} {\bibfnamefont{P.}~\bibnamefont{Davidson}}, \bibinfo {author}
  {\bibfnamefont{J.}~\bibnamefont{Ferr\'e}}, \bibinfo {author}
  {\bibfnamefont{J.~P.}\ \bibnamefont{Jamet}}, \bibinfo {author}
  {\bibfnamefont{P.}~\bibnamefont{Panine}}, \bibinfo {author}
  {\bibfnamefont{I.}~\bibnamefont{Dozov}},\ and\ \bibinfo {author}
  {\bibfnamefont{J.~P.}\ \bibnamefont{Jolivet}},\ }%
  \bibfield{journal}{%
  \bibinfo {journal} {Physical Review Letters}\ }%
  \textbf{\bibinfo {volume} {88}},\ \bibinfo {pages} {125507} (\bibinfo {year}
  {2002})\BibitemShut{NoStop}%
\bibitem{Fluerasu07}%
  \BibitemOpen
  \bibfield{author}{%
  \bibinfo {author} {\bibfnamefont{A.}~\bibnamefont{Fluerasu}}, \bibinfo
  {author} {\bibfnamefont{A.}~\bibnamefont{Moussa\"{\i}d}}, \bibinfo {author}
  {\bibfnamefont{A.}~\bibnamefont{Madsen}},\ and\ \bibinfo {author}
  {\bibfnamefont{A.}~\bibnamefont{Schofield}},\ }%
  \bibfield{journal}{%
  \bibinfo {journal} {Physical Review E}\ }%
  \textbf{\bibinfo {volume} {76}},\ \bibinfo {pages} {010401(R)} (\bibinfo
  {year} {2007})\BibitemShut{NoStop}%
\bibitem{Naegele97}%
  \BibitemOpen
  \bibfield{author}{%
  \bibinfo {author} {\bibfnamefont{G.}~\bibnamefont{N\"{a}gele}}\ and\ \bibinfo
  {author} {\bibfnamefont{P.}~\bibnamefont{Baur}},\ }%
  \bibfield{journal}{%
  \bibinfo {journal} {Physica A}\ }%
  \textbf{\bibinfo {volume} {245}},\ \bibinfo {pages} {297} (\bibinfo {year}
  {1997})\BibitemShut{NoStop}%
\bibitem{Solomon98}%
  \BibitemOpen
  \bibfield{author}{%
  \bibinfo {author} {\bibfnamefont{M.~J.}\ \bibnamefont{Solomon}}\ and\
  \bibinfo {author} {\bibfnamefont{D.~V.}\ \bibnamefont{Boger}},\ }%
  \bibfield{journal}{%
  \bibinfo {journal} {Journal of Rheology}\ }%
  \textbf{\bibinfo {volume} {42}},\ \bibinfo {pages} {929} (\bibinfo {year}
  {1998})\BibitemShut{NoStop}%
\bibitem{Tirado84}%
  \BibitemOpen
  \bibfield{author}{%
  \bibinfo {author} {\bibfnamefont{M.~M.}\ \bibnamefont{Tirado}}, \bibinfo
  {author} {\bibfnamefont{C.~L.}\ \bibnamefont{Martinez}},\ and\ \bibinfo
  {author} {\bibfnamefont{J.~G.}\ \bibnamefont{de~la Torre}},\ }%
  \bibfield{journal}{%
  \bibinfo {journal} {The Journal of Chemical Physics}\ }%
  \textbf{\bibinfo {volume} {81}},\ \bibinfo {pages} {2047} (\bibinfo {year}
  {1984})\BibitemShut{NoStop}%
\bibitem{Beenakker84}%
  \BibitemOpen
  \bibfield{author}{%
  \bibinfo {author} {\bibfnamefont{C.~W.~J.}\ \bibnamefont{Beenakker}}\ and\
  \bibinfo {author} {\bibfnamefont{P.}~\bibnamefont{Mazur}},\ }%
  \bibfield{journal}{%
  \bibinfo {journal} {Physica A}\ }%
  \textbf{\bibinfo {volume} {120}},\ \bibinfo {pages} {388} (\bibinfo {year}
  {1984})\BibitemShut{NoStop}%
\bibitem{Note1}%
  \BibitemOpen
  \bibinfo {note} {Considering an ionic strength of 10~mM, the data of Solomon
  and Boger \cite {Solomon98} give a value of $\eta _0 / \eta = 0.41$ using the
  low concentration expansion and 0.25 using the overall formula (for an aspect
  ratio of 8.4). Rheology measurements performed at room temperature in aqueous
  suspension at the same concentration (6.7~\%) also yield $\eta _0 / \eta =
  0.25$, but the comparison is not straightforward due to the lack of
  information on the electrostatic effects in the presence of
  propane-1,3-diol.}\BibitemShut{Stop}%
\bibitem{Banchio06}%
  \BibitemOpen
  \bibfield{author}{%
  \bibinfo {author} {\bibfnamefont{A.~J.}\ \bibnamefont{Banchio}}, \bibinfo
  {author} {\bibfnamefont{J.}~\bibnamefont{Gapinski}}, \bibinfo {author}
  {\bibfnamefont{A.}~\bibnamefont{Patkowski}}, \bibinfo {author}
  {\bibfnamefont{W.}~\bibnamefont{H\"{a}u{\ss}ler}}, \bibinfo {author}
  {\bibfnamefont{A.}~\bibnamefont{Fluerasu}}, \bibinfo {author}
  {\bibfnamefont{S.}~\bibnamefont{Sacanna}}, \bibinfo {author}
  {\bibfnamefont{P.}~\bibnamefont{Holmqvist}}, \bibinfo {author}
  {\bibfnamefont{G.}~\bibnamefont{Meier}}, \bibinfo {author}
  {\bibfnamefont{M.~P.}\ \bibnamefont{Lettinga}},\ and\ \bibinfo {author}
  {\bibfnamefont{G.}~\bibnamefont{N\"{a}gele}},\ }%
  \bibfield{journal}{%
  \bibinfo {journal} {Physical Review Letters}\ }%
  \textbf{\bibinfo {volume} {96}},\ \bibinfo {pages} {138303} (\bibinfo {year}
  {2006})\BibitemShut{NoStop}%
\bibitem{Note2}%
  \BibitemOpen
  \bibinfo {note} {We define a 2D system as invariant under translation along
  the normal direction, in contrast with the extensively studied quasi-2D
  systems that consist of particles confined at an interface or between rigid
  boundaries.}\BibitemShut{Stop}%
\bibitem{Happel83}%
  \BibitemOpen
  \bibfield{author}{%
  \bibinfo {author} {\bibfnamefont{J.}~\bibnamefont{Happel}}\ and\ \bibinfo
  {author} {\bibfnamefont{H.}~\bibnamefont{Brenner}},\ }%
  \emph{\bibinfo {title} {Low {Reynolds} number hydrodynamics}}\ (\bibinfo
  {publisher} {Martinus Nijhoff},\ \bibinfo {year} {1983})\BibitemShut{NoStop}%
\bibitem{Falck04}%
  \BibitemOpen
  \bibfield{author}{%
  \bibinfo {author} {\bibfnamefont{E.}~\bibnamefont{Falck}}, \bibinfo {author}
  {\bibfnamefont{J.~M.}\ \bibnamefont{Lahtinen}}, \bibinfo {author}
  {\bibfnamefont{I.}~\bibnamefont{Vattulainen}},\ and\ \bibinfo {author}
  {\bibfnamefont{T.}~\bibnamefont{Ala-Nissila}},\ }%
  \bibfield{journal}{%
  \bibinfo {journal} {The European Physical Journal E-Soft Matter}\ }%
  \textbf{\bibinfo {volume} {13}},\ \bibinfo {pages} {267} (\bibinfo {year}
  {2004})\BibitemShut{NoStop}%
\end{thebibliography}
\end{document}